# A Comparison of Source Distribution and Result Overlap in Web Search Engines

**AUTHORS SECTION**

**Yagci, Nurce**  HAW Hamburg, Germany | nurce.yagci@haw-hamburg.de

**Sünkler, Sebastian**  HAW Hamburg, Germany | sebastian.suenkler@haw-hamburg.de

**Häußler, Helena**  HAW Hamburg, Germany | helena.haeuessler@haw-hamburg.de

**Lewandowski, Dirk**  HAW Hamburg, Germany | dirk.lewandowski@haw-hamburg.de

**ABSTRACT**

When it comes to search engines, users generally prefer Google. Our study aims to find the differences between the results found in Google compared to other search engines. We compared the top 10 results from Google, Bing, DuckDuckGo, and Metager, using 3,537 queries generated from Google Trends from Germany and the US. Google displays more unique domains in the top results than its competitors. Wikipedia and news websites are the most popular sources overall. With some top sources dominating search results, the distribution of domains is also consistent across all search engines. The overlap between Google and Bing is always under 32%, while Metager has a higher overlap with Bing than DuckDuckGo, going up to 78%. This study shows that the use of another search engine, especially in addition to Google, provides a wider variety in sources and might lead the user to find new perspectives.

**INTRODUCTION**

Why should there be more than one search engine? While users may prefer one search engine over others for its usability, specialized features, or a more convenient integration into their technical environment, the question that interests us in this research is whether a user will benefit from using another search engine than Google when it comes to finding results from different sources. Our starting point is the fact that Google is the most-used search engine by far (StatCounter, 2022), that user to a large degree trust search engines to provide them relevant and useful results (European Commission, 2016; Purcell et al., 2012), and that only some users use another search engine in addition to Google (Schultheiß & Lewandowski, 2021).

Users place great trust in search engines. This is reflected by the 91% of US users who said they find what they are looking for always or most of the time, and the 66% who believe search engines are a fair and unbiased source of information (Purcell et al., 2012). Furthermore, 78% of European internet and online platform users said they trust that their search engine results are the most relevant results (European Commission, 2016). Globally, users trust search engines more than any other source (including traditional news outlets) when it comes to news (Edelman Trust Institute, 2022) and users trust news found via search significantly more than news found on social media (Newman et al., 2021).

As the Web is enormous and different search engines might prefer different sources, it is interesting to see whether the top sources shown in search results differ from one search engine to the other. It might be that an alternative search engine prefers results from "alternative" sources, e.g., in terms of political leaning or preferring non-commercial content providers. This all comes down to whether alternative search engines are actually alternatives in regards to the results they display. If they were, possible benefits of using a search engine other than Google include finding different results, finding additional results, and finding more relevant results. No matter which of these goals a user aims to achieve, they will need *other* results than Google's. Therefore, it is interesting to see whether other search engines provide users with such results.

There has been an ongoing discussion on alternative search engines and how Google's dominance in the search engine market can be broken. Approaches range from establishing single alternative search engines to building infrastructures for such alternatives (e.g., Lewandowski, 2019); also see Mager, 2014). With Google dominating the search engine market (StatCounter, 2022), it often seems that there are no alternatives at all. On the other hand, the number of alternative (or simply "other") search engines is often overestimated. Many seem-to-be search engines are merely search portals displaying results from a partner instead of generating the results from their own index. For instance, Yahoo and Ecosia get their results from Bing and can therefore not be considered search engines in their own right. But still, there may be other reasons for using a search engine without its own index. Some of the unique benefits alternative search engines advertise are privacy (e.g., Startpage and DuckDuckGo) or being a company investing its profits in environmental projects (e.g., Ecosia). Another type of search engine is the meta search engine

(e.g., Metager). Such an engine sends the queries to several other search engines, then aggregates and re-ranks the top results. We deem it especially interesting whether such an approach will lead to a wider variety of search results, i.e., results from a more diverse set of sources. So, in the context of our research, we will consider any search engine that either has its own index or provides a unique selection and re-ranking of results from one or more indexes as an alternative search engine. We are especially interested in the differences in the source distribution; the relevance of the results is out of the scope of our research.

More than 20 years ago, Introna & Nissenbaum (2000) argued that search engines as commercial operations tend to prefer big websites and, therefore, a portion of the Web, i.e., the smaller sites, remain hidden from view. Studies measuring what users select seem to confirm this: Goel et al. (2010) found that within Yahoo, only 10,000 websites account for approximately 80% of result clicks. It is important to note that this does not merely result from user preference for particular sources but that users predominantly select from the top results shown by the search engine. What is out of the immediate view of users will not be chosen (Lewandowski & Kammerer, 2021).

It is striking that few studies have compared the results between different search engines in recent years. Older studies (see literature review section) overall found that top results from different search engines did not overlap too much. In this paper, we address how the top results of Google differ from alternatives and, therefore, whether it is worthwhile for a user to consider these alternatives. If a search engine other than Google produced very similar results to Google, a user would not benefit much from using that search engine when source variety is considered.

## LITERATURE REVIEW

### Search result overlap

By the mid-1990s, researching the overlap of search results between different search engines sparked interest for the purpose of estimating the size of the Web (Bharat, 1998; Ding & Marchionini, 1996; Lawrence & Giles, 1999). The generally small degrees of overlap indicated diverse underlying databases, each of limited size. Therefore, meta-search engines that combined the results of several search engines were meant to provide an additional value (Chignell et al., 1999) and attracted further research (i.e., Meng et al., 2002). Consequently, Spink et al. (Spink et al., 2006) included the meta search engine Dogpile in their extensive study on the overlap of search results. For a collection of 12,570 queries, the results found on the first search engine result page (SERP) of Ask Jeeves, Google, MSN Search, and Yahoo were captured. 84.9% of the results were unique to one search engine, while only 1.1% were shared by all search engines. Additionally, the low overlap between search engines also manifests in the ranking since only 7% of the top results were similar. This is consistent with previous studies that reported low overlap in ranking (Bar-Ilan, 2005; Bar-Ilan et al., 2006).

Subsequent studies report an increase of overlap in search results, while ranking algorithms appear to be the leading cause for differences in the presentation of the results. Bilal and Ellis (2011) compared major web search engines (Google, Bing, Yahoo) with search engines specifically designed for children (Ask Kids, Yahoo Kids) for queries of various lengths. Bing and Yahoo shared the same overlap with Google for nearly all queries, ranging between 22% and 40%. In contrast, Yahoo Kids had mostly unique results. However, the disparate relevance ranking between Bing and Yahoo suggests different ranking algorithms. Cardoso and Magalhães (2011) measured the overlap of search result rankings based on URLs and website contents. Search results for 40,000 queries were retrieved from Google and Bing. The overlap in domains was about 29%, and Google had more exclusive domains. When looking at the result sets without considering the positions, the similarity between the result sets increased. This indicates that Google and Bing have different ranking preferences but index mostly the same sources.

Similarly, Agrawal et al. (2016) based their overlap analysis on content but factored in the description of the result snippets. The top 10 results of 67 informational queries were collected from Google and Bing. The results indicate a high overlap between Google and Bing, with a slightly higher similarity among the top 5 results. Most recently, a study about information on Covid-19 was conducted by Makhortykh et al. (2020). The first 50 results for queries in different languages were collected from Baidu, Bing, DuckDuckGo, Google, Yandex, and Yahoo. DuckDuckGo and Yahoo shared nearly 50% of their results, whereas the other pairs remained under 25%. Google had an overlap of around 10% with Bing and a negligible one with DuckDuckGo, Yandex, Yahoo, and Baidu.

### Source diversity

Various studies took a closer look at search results and evaluated the domains, the types of sources, as well as their diversity across result sets and search engines. Comparing Google, Live Search, and Yahoo, Thelwall (2008) reported that Yahoo returned the highest amount of different domains to a query and also returned around 10% more top-level domains than other search engines. Höchstötter & Lewandowski (2009) showed that while the sources in top results of various search engines differ, there is a concentration on some popular sources in top results. More recently, Lewandowski & Sünkler (2019) collected search results on queries related to insurance providers and identified the distribution of top-level domains. Only ten different domains were found in the first result across all queries. The five most popular domains in the top 5 results are price comparison websites, which make up 88.4% of all first results. The share of popular domains decreases according to the ranking but remains at 42.9% at position ten.

Unkel and Haim (2019) observed Google search results prior to the German parliamentary elections in 2017. The study found that result lists for candidates and parties exhibit a high share of self-administered websites. In contrast, results about election facts, guidance, and issues are mainly from general interest news, government information, and privately run websites. Among the top ten domains across all search results are seven news websites which account for a quarter of all search results. By contrast, Wikipedia made up 5.4% of all search results. The high share of news and Wikipedia articles is confirmed by Steiner et al. (2022) who compared the first results for queries on debated topics in Germany across the search engines Ask, Bing, DuckDuckGo, Google, and Ixquick. All search engines positioned Wikipedia in the first rank for queries on climate change and the Transatlantic Trade and Investment Partnership (TTIP); Ask did so for almost every query. In the other search engines, news sites were placed in the first rank most of the time. However, for the topic Covid-19, the sources are more diverse. As Makhortykh et al. (2020) observed, only Yahoo incorporates recent information from legacy media, whereas Bing strongly relied on healthcare-related sources and Google highlighted government-related websites. Yandex was the only search engine that included alternative media in its top 20 search results. Especially for recent topics that lack an established knowledge basis, the choice of search engine may considerably impact what information a user gets to see.

## OBJECTIVES AND RESEARCH QUESTIONS
Our study addresses the following research questions:

1. Do top results from alternative search engines differ from Google's in regard to the number of unique sources?

2. Do top results from alternative search engines differ from Google's in regard to top sources?

3. Do top results from alternative search engines differ from Google's in regard to source concentration, i.e., are results distributed over more or fewer sources in different search engines?

To answer these questions, we selected three alternative search engines to compare to Google. Aside from Bing, which is the biggest competitor to Google, we chose DuckDuckGo and Metager. DuckDuckGo uses results from Bing, Yahoo, and Yandex, and has the added benefit of advanced privacy settings such as non-personalized results. Metager, a German meta-search engine, aggregates results from several search engines, including Bing and Yandex. As such, either search engine might provide a more varied set of sources. Since the alternative search engines each have an index different from Google's, the comparison will give insights into what domains are favored by Google and which domains are excluded. Also, we will gain insights into the differences in ranking between them.

## METHODS
As illustrated in Figure 1, this study was conducted in five steps. In the first step, the query sets were generated from previously collected Google Trends data. The daily trends for both Germany and the US were collected daily at 3 am CET from November 10$^{th,}$ 2021, until March 31$^{st,}$ 2022. Since the daily number of trending queries can vary, the query sets for Germany and the US are not the same size. Two thousand nine hundred sixteen trending queries for Germany and 2,819 for the US were initially collected. However, since some topics of interest are longer lived than a single day, the same topics can be trending on multiple days. Therefore, after removing duplicates, our query sets consist of 1,821 queries for Germany and 2,126 queries for the US. The queries have a wide range, including topics like celebrities (e.g., "Sandra Bullock"), sports (e.g., "NFL," "Liverpool vs. Southampton"), seasonal events (e.g., "Macy's Thanksgiving Day Parade," "Restaurants open on Thanksgiving"), headline news (e.g., "TikTok class-action lawsuit settlement"), and more.

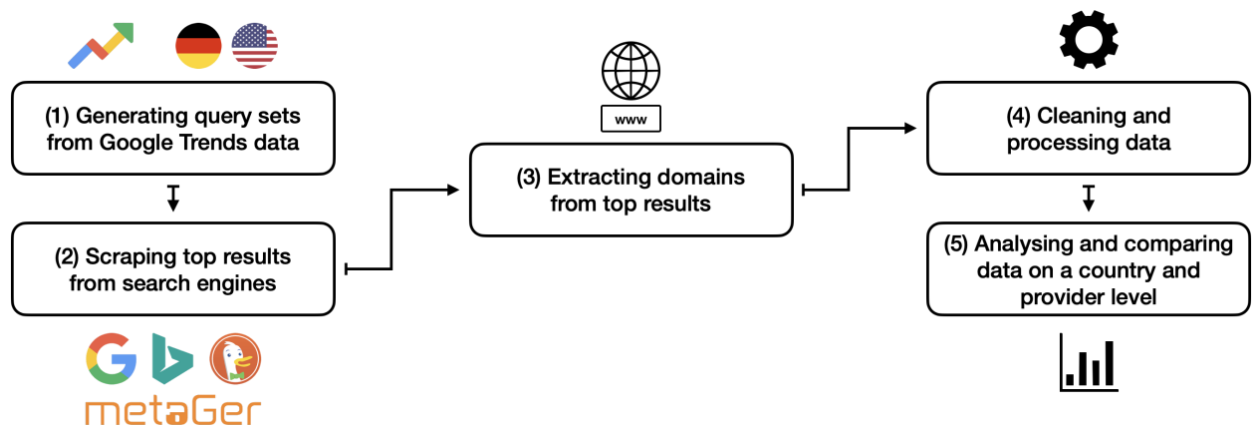

**Figure 1. Methodology of the current study**

In step 2, for each query, the top 10 results from all four search engines were collected. We made sure to request results for the appropriate language and location by using URL parameters. Furthermore, we decided to limit the comparison to the top 10 results because users usually only consider these. Again, for our study, only the implications for the user perception are relevant. The Web scraping component of the Relevance Assessment Tool (RAT; Sünkler et al., 2022) was used to collect the data. The scraping took place between April 6$^{th}$ and 9$^{th,}$ 2022, with 230,315 results (Germany: 109,604; US: 120,711) gathered. It is important to note that only organic search results were considered. A fair comparison can be ensured when ranked result lists instead of vertical inserts of universal search results (e.g., Google News and Bing News) are considered. We also ignored ads since they are not part of the search engine's web index.

During step 3, the python library urllib was used to get the domain of each search result. To ensure a fair comparison between all search engines, in step 4, we removed results for any query that had less than ten results in any of the search engines. Additionally, the data was checked and stripped of any errors that might have occurred during the collection, like duplicate results. This resulted in a further reduced dataset consisting of 1,672 queries and 66,880 results for Germany and 1,865 queries and 74,600 results for the US. Lastly, the extraction of top-level domains was further refined in this step. We used basic string matching to unify URLs that pointed to the same domains. For example, the following three URLs would have been counted as separate domains without the string matching: https://www.facebook.com, "http://www.facebook.com", and "http://www.facebook.com/".

Finally, in step five, we used different methods of comparing the data collected. An initial comparison is made by using simple descriptive statistics. To measure source concentration, we adapt the Gini coefficient, a measure of statistical dispersion used to measure income or wealth inequality (Gini, 1936). The Gini coefficient is a single number ranging from 0 to 1, where 0 represents perfect equality, and 1 is the maximum inequality. As Ortega et al. (2008) demonstrated, adapting this measure to the distribution of values in different domains is appropriate. In the case of search result sets, this allows us to compare the various search engines easily. The lower the Gini coefficient, the more equally the results are distributed over all domains contributing to a particular result set.

Following that, we calculate the Jaccard similarity index (González et al., 2008). Puschmann (2019) demonstrates the usage of the Jaccard index to measure the similarity between two result sets. The Jaccard index is a single number ranging from 0 to 1, where 0 means that the two sets are entirely dissimilar, while 1 means that they are identical. We calculate the Jaccard index for every two-way combination of our four search engines (e.g., Google and Bing, Bing and Metager, etc.) for all stages between the top 1 and top 10 results (e.g., Google and Bing position 1 to 5).

## RESULTS

### Classification of domains

Since we are working with sources from two different countries and the type of content behind some domains might not be immediately recognizable, we manually created a general set of categories based on the top 50 domains found across all results. Following this, we manually classified the 50 most popular domains for both the German and US results (see Table 1).

When comparing the top 50 domains in the German results, News service dominates with 54%, while Movies & Entertainment and Sports make up 18% and 14%, respectively. In the US, 34% of the top 50 domains are Sports websites, with News services following close behind at 30%. Movies & Entertainment only make up 12%. These are the overall distributions of domain classes across all search engines; below, we look at the differences *between* the search engines.

| Class | Example | Top 50 Germany (share) | Top 50 US (share) |
|---|---|---|---|
| Celebrities | variety.com | 0.06 | 0.10 |
| E-Commerce | amazon.com | 0.02 | 0.06 |
| Government website | bundesregierung.de | 0 | 0 |
| Information service | wikipedia.org | 0.02 | 0.02 |
| Movies & Entertainment | imdb.com | 0.18 | 0.12 |
| News service | cnn.com | 0.54 | 0.30 |
| Sports | espn.com | 0.14 | 0.34 |
| Social media | instagram.com | 0.04 | 0.06 |

**Table 1. Classes of domains and their frequency in the top 50 domains**

**Variety of domains**

A comparison of the number of root domains in the search results of Google, Bing, DuckDuckGo, and Metager in Germany shows that Google has the greatest diversity by a small margin (see Figure 2). Overall, the values are very similar. However, it is noticeable that Google has the greatest variety of domains, especially in the first three positions. Interestingly, the greater diversity of Google's sources is even more pronounced in the US results. To examine the differences more closely, we look at the numbers below.

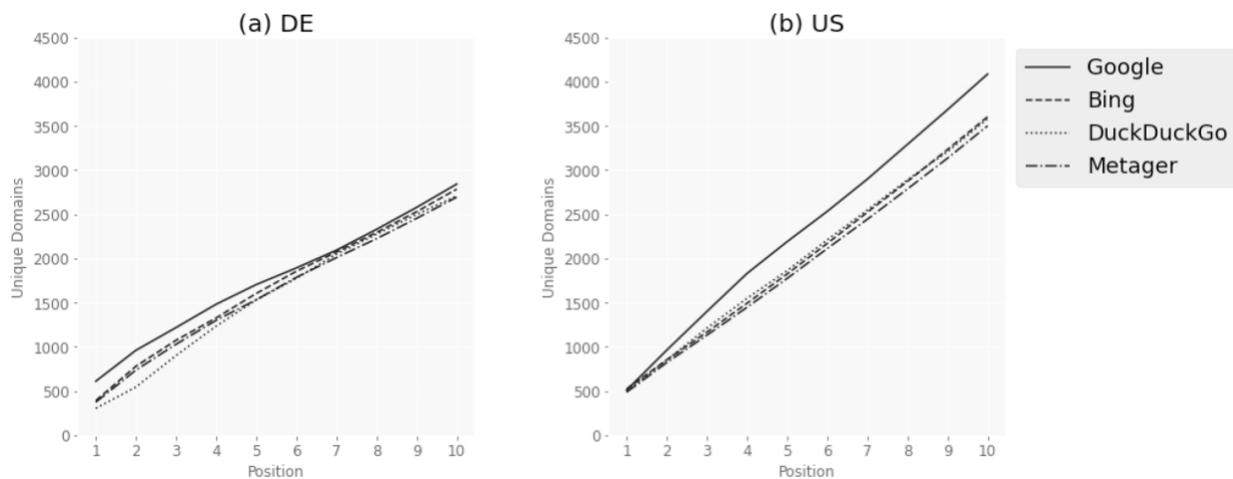

**Figure 2. Cumulative number of unique domains**

Table 2 shows the cumulative frequencies for the German results. The number of unique root domains converges at the fifth position. We found 2,841 unique domains in Google's results, 2,783 unique domains for Bing, 2,707 unique domains for DuckDuckGo, and 2,683 unique domains for Metager in Germany.

| Position | Google | | Bing | | DuckDuckGo | | Metager | |
|---|---|---|---|---|---|---|---|---|
| | # | share | # | share | # | share | # | share |
| 1 | 609 | 0.21 | 389 | 0.14 | 302 | 0.11 | 372 | 0.14 |
| 2 | 959 | 0.34 | 778 | 0.28 | 544 | 0.20 | 735 | 0.27 |
| 3 | 1,216 | 0.43 | 1,072 | 0.39 | 900 | 0.33 | 1,027 | 0.38 |
| 4 | 1,480 | 0.52 | 1,325 | 0.48 | 1,232 | 0.46 | 1,298 | 0.48 |
| 5 | 1,703 | 0.60 | 1,602 | 0.58 | 1,530 | 0.57 | 1,531 | 0.57 |

| Position | Google | | Bing | | DuckDuckGo | | Metager | |
|---|---|---|---|---|---|---|---|---|
| | # | share | # | share | # | share | # | share |
| 6 | 1,891 | 0.67 | 1,853 | 0.67 | 1,773 | 0.65 | 1,785 | 0.66 |
| 7 | 2,092 | 0.74 | 2,071 | 0.74 | 2,047 | 0.76 | 2,008 | 0.75 |
| 8 | 2,329 | 0.82 | 2,292 | 0.82 | 2,272 | 0.84 | 2,222 | 0.83 |
| 9 | 2,577 | 0.91 | 2,528 | 0.91 | 2,497 | 0.92 | 2,453 | 0.91 |
| 10 | 2,841 | 1.00 | 2,783 | 1.00 | 2,707 | 1.00 | 2,693 | 1.00 |

Table 2. Cumulative number of unique domains (Germany; #: cumulative number, %: percentage of total)

There are differences when comparing the search results in Germany and the US. It is notable that, in contrast to Germany, in the US results, there are only minor differences in the first positions. However, Google also shows the greatest variety of root domains in the search results in the US. This is more evident than in the German results (see Table 3). Overall, we found 4,085 unique domains in Google, 3,602 in Bing, 3,579 in DuckDuckGo, and 3,500 in Metager.

| Position | Google | | Bing | | DuckDuckGo | | Metager | |
|---|---|---|---|---|---|---|---|---|
| | # | share | # | share | # | share | # | share |
| 1 | 508 | 0.12 | 521 | 0.14 | 493 | 0.14 | 485 | 0.14 |
| 2 | 961 | 0.24 | 851 | 0.24 | 849 | 0.24 | 820 | 0.23 |
| 3 | 1,398 | 0.34 | 1,164 | 0.32 | 1,210 | 0.34 | 1,130 | 0.32 |
| 4 | 1,828 | 0.45 | 1,491 | 0.41 | 1,547 | 0.43 | 1,445 | 0.41 |
| 5 | 2,188 | 0.54 | 1,822 | 0.51 | 1,860 | 0.52 | 1,772 | 0.51 |
| 6 | 2,531 | 0.62 | 2,168 | 0.60 | 2,208 | 0.62 | 2,110 | 0.60 |
| 7 | 2,896 | 0.71 | 2,521 | 0.70 | 2,546 | 0.71 | 2,442 | 0.70 |
| 8 | 3,289 | 0.81 | 2,870 | 0.80 | 2,888 | 0.81 | 2,785 | 0.80 |
| 9 | 3,682 | 0.90 | 3,231 | 0.90 | 3,207 | 0.90 | 3,133 | 0.90 |
| 10 | 4,085 | 1.00 | 3,602 | 1.00 | 3,579 | 1.00 | 3,500 | 1.00 |

Table 3. Cumulative number of unique domains (USA; #: cumulative number, %: percentage of total)

**Popular domains**
When comparing the top domains for each search engine across all German search results collected, we find a clear preference for Wikipedia in all of them (see Table 4). While Wikipedia is the most popular domain in all search engines, the frequency in the Google results is significantly lower, with only 658 compared to the number of occurrences in Bing being 1,948, in Metager 1,878, and in DuckDuckGo 1,752.

The other domains in the top 10 across the search engines are News services (Google: 7, Bing: 4, DuckDuckGo: 3, Metager: 4). Sports make up 2 out of 10 for all but Google, with only one sports website. Surprisingly, Instagram is in the top 10 for DuckDuckGo and Metager, even though in Table 1, we showed that domains of the class Social Media only make up 4% of top domains in the German results. The same goes for Amazon, which is the second most frequent domain in DuckDuckGo, even though in the overall results, E-Commerce websites only make up 2%.

| No. | Google | Bing | DuckDuckGo | Metager |
|---|---|---|---|---|
| 1 | wikipedia.org (658) | wikipedia.org (1,948) | wikipedia.org (1,752) | wikipedia.org (1,878) |
| 2 | spiegel.de (414) | zdf.de (324) | amazon.de (735) | zdf.de (344) |
| 3 | stern.de (375) | bild.de (298) | zdf.de (298) | bunte.de (314) |

| No. | Google | Bing | DuckDuckGo | Metager |
|---|---|---|---|---|
| 4 | gala.de (300) | bunte.de (296) | kicker.de (261) | bild.de (290) |
| 5 | bild.de (288) | kicker.de (294) | bild.de (255) | kicker.de (285) |
| 6 | t-online.de (279) | gala.de (279) | tagesschau.de (251) | gala.de (282) |
| 7 | tagesschau.de (271) | tagesschau.de (274) | wunderweib.de (248) | tagesschau.de (268) |
| 8 | kicker.de (222) | sportschau.de (261) | bunte.de (245) | sportschau.de (256) |
| 9 | zdf.de (219) | web.de (252) | instagram.com (243) | web.de (238) |
| 10 | sueddeutsche.de (219) | instagram.com (245) | sportschau.de (242) | instagram.com (232) |

**Table 4. Top 10 domains shown in different search engines (Germany)**

Contrarily, in the US results, Wikipedia has the highest number of occurrences in Google with 1,892 compared to Bing's 1,388, Metager's 1,304, and DuckDuckGo's 1,287 (see Table 5). Still, what remains the same is that Wikipedia is the most popular domain across all search engines.

Another difference in the US results is that Social Media sources are much more prevalent across all search engines, especially in Google, with Instagram as the second most frequent domain, Facebook as the third, and Twitter as the sixth. This is surprising when considering that Social Media domains only make up 6% of the top domains for the US results (see Table 1). Another finding is that YouTube, a subsidiary of Google, is the top 8th domain for all three alternatives but not for Google. Again, 30% of the top domains for Google, DuckDuckGo, and Metager are News services. For Bing, it is 40%. For each search engine, Sports websites make up 20% of domains.

| No. | Google | Bing | DuckDuckGo | Metager |
|---|---|---|---|---|
| 1 | wikipedia.org (1,892) | wikipedia.org (1,388) | wikipedia.org (1,287) | wikipedia.org (1,304) |
| 2 | instagram.com (726) | imdb.com (669) | imdb.com (644) | yahoo.com (714) |
| 3 | facebook.com (503) | yahoo.com (634) | yahoo.com (642) | imdb.com (680) |
| 4 | espn.com (377) | espn.com (610) | espn.com (601) | espn.com (614) |
| 5 | cbssports.com (363) | nypost.com (518) | nypost.com (536) | nypost.com (514) |
| 6 | twitter.com (328) | cbssports.com (511) | cbssports.com (513) | cnn.com (504) |
| 7 | imdb.com (244) | cnn.com (495) | cnn.com (506) | cbssports.com (502) |
| 8 | britannica.com (226) | youtube.com (414) | youtube.com (352) | youtube.com (417) |
| 9 | usatoday.com (222) | msn.com (322) | go.com (320) | msn.com (324) |
| 10 | nytimes.com (220) | instagram.com (318) | instagram.com (281) | instagram.com (288) |

**Table 5. Top 10 domains for each search engine (USA)**

**Exclusivity of domains**
This section compares the top 50 domains for each search engine to determine what source Google and its alternatives might exclusively offer to the user. Table 6 shows the domains found solely in Google's top 50 domains list for the German queries. Out of these 20 domains, eight are Sports websites, six are News services, and four belong to the Entertainment & Movies class, with one each for E-Commerce and Government websites.

| ran.de | sky.de | dfb.de | spox.com |
|---|---|---|---|
| amazon.de | eurosport.de | kino.de | sportbuzzer.de |
| dazn.com | zeit.de | fr.de | deutschlandfunk.de |
| fernsehserien.de | tagesspiegel.de | filmstarts.de | rnd.de |
| bundesregierung.de | tz.de | weltfussball.de | dw.com |

On the contrary, table 7 shows the list of domains found in the top 50 domains of all three alternatives but not in Google. These would be the domains missed by users who only use Google. Out of the 13 domains, four are News services, followed by three Movies & Entertainment and two Celebrities sources. Interestingly, Instagram and YouTube are missing from Google, as well.

| sport.de | instagram.com | finanzen.net | ardmediathek.de |
|---|---|---|---|
| wunderweib.de | fussballdaten.de | youtube.com | express.de |
| promipool.de | n-tv.de | imdb.com | daserste.de |
| abendzeitung-muenchen.de | | | |

Table 7. Domains found in all search engines but Google's top 50 (Germany)

When implementing the same evaluation for the US results, we find that 17 domains are exclusively found in Google's top 50 domains, with five domains classified as Sports, four as Movies & Entertainment, another four as News services, three as Social Media and one as Celebrities (see Table 8). Notable is that social media giants Facebook and TikTok are only found in Google and not in the alternatives.

| marca.com | reuters.com | rotowire.com | facebook.com | tiktok.com |
|---|---|---|---|---|
| usmagazine.com | npr.org | discogs.com | spotify.com | theathletic.com |
| spotrac.com | linkedin.com | deadline.com | forbes.com | variety.com |
| washingtonpost.com | sports-reference.com | | | |

Table 8. Domains only found in Google's top 50 (US)

On the opposite end, 13 domains are not found in the US results on Google (see Table 9). Six of them are News services, three are websites about Celebrities, two are about Movies & Entertainment, and one each are Sports and E-Commerce websites. Interestingly, Amazon is missing in Google's results, as well as msn.com, a service from Microsoft, like Bing itself.

| amazon.com | msn.com | heavy.com | tmz.com |
|---|---|---|---|
| yardbarker.com | cnn.com | foxnews.com | newsweek.com |
| pagesix.com | screenrant.com | decider.com | huffpost.com |
| biography.com | | | |

Table 9. Domains found in all search engines but Google's top 50 (US)

**Distribution of domains**
Next, we look at the Gini coefficient to measure the concentration and statistical dispersion of root domains in search engines. The distributions across all search engines in both German and US results show very similar values. The results range from 0.77 to 0.79 and 0.73 to 0.76 for German and US results, respectively (see Figure 3). Overall, this means that there is an imbalance in the distribution of results from root domains and that some top sources dominate the search results for both countries.

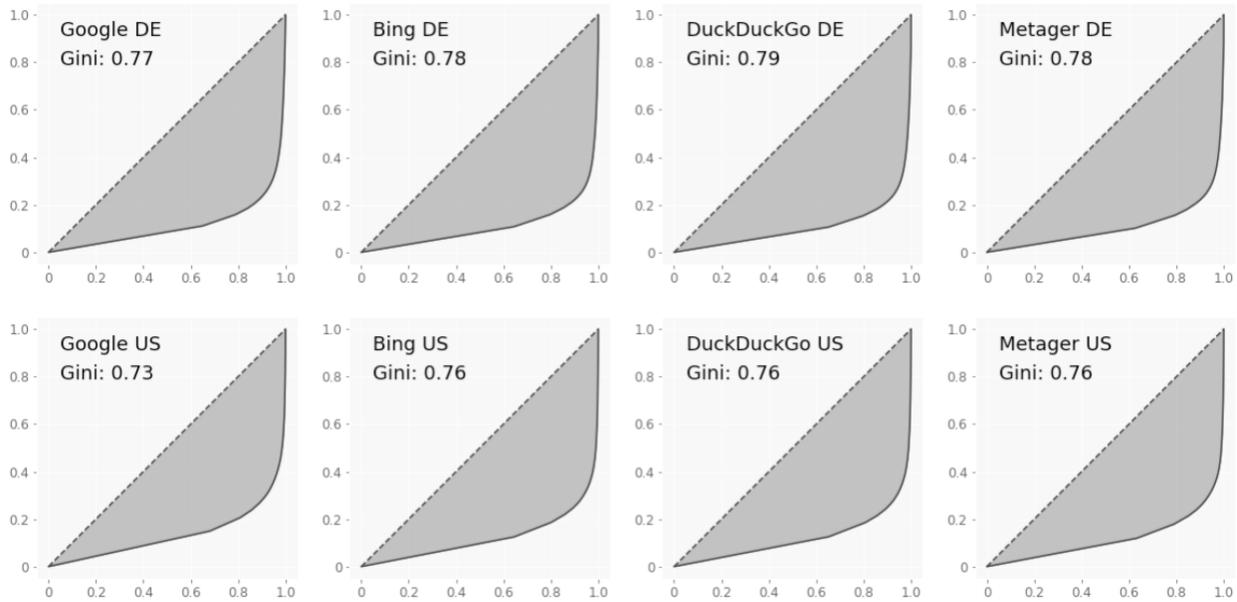

**Figure 3. Source distribution of domains (Germany)**

**Similarity of domains**

Comparing the similarities of the top 10 German results, we find that Bing and Metager are most similar in terms of their top results, with 70%, followed by 64% overlap between DuckDuckGo and Metager and 63% between Bing and DuckDuckGo. However, when comparing the alternative search engines to Google, we find that the highest overlap is between Google and Bing with 28% and slightly lower ones with DuckDuckGo and Metager at 27%.

Interestingly, the top 1 result overlap is higher between Google and Bing (31%) than the overlap between Bing and DuckDuckGo (15%).

The results show a mean overlap of 30% of top 1, 40% of top 3, and 47% of top 10 results (table 10). This indicates that when looking at the entire first search results page, nearly half of the domains are the same in all search engines considered. For search engine users, this means that in addition to Google, this would also allow them to see search results that they would otherwise miss if they used the alternative search engine

| Rank | Google Bing | Google DDG | Google Metager | Bing DDG | Bing Metager | DDG Metager | Mean |
|---|---|---|---|---|---|---|---|
| Top 1 | 0.31 | 0.10 | 0.29 | 0.15 | 0.78 | 0.15 | 0.30 |
| Top 3 | 0.30 | 0.21 | 0.30 | 0.39 | 0.78 | 0.39 | 0.40 |
| Top 10 | 0.28 | 0.27 | 0.27 | 0.63 | 0.70 | 0.64 | 0.47 |

**Table 10. Jaccard similarity of domains (Germany; *DDG: DuckDuckGo*)**

The same evaluation for the US yields similar results. Again, the overlap between Bing and Metager is the highest, but the margin between this pair and the other alternative pairings has been reduced. When looking at the top 10 results, Bing and Metager overlap by 65%, Bing and DuckDuckGo by 64%, and DuckDuckGo and Metager by 62%. While Google's overlap with the other search engines is lower in the US results, again, the pair of Google and Bing is slightly higher (25%) than the 24% overlap between Google and the other two alternatives (see Table 11).

Narrowing the results down to the top 3, the overlap between Bing and DuckDuckGo increases minimally to 65% and remains the same for DuckDuckGo and Metager at 62%. Here, the overlap between Google and Bing is lower than in the German results, with only 26%. Furthermore, the top 1 result overlap between Google and Bing is at 29%, which is also lower than in the German results. Generally, compared to the German results, there is a higher overlap in the top 1 position for the pairs Google and DuckDuckGo (20%), Bing and DuckDuckGo (50%), and DuckDuckGo and Metager (47%). The averaged results make this trend clearer because the overlaps for the top 1, top 3, and top 10 results are similar, with 41%, 46%, and 44%, respectively.

| Rank | Google Bing | Google DDG | Google Metager | Bing DDG | Bing Metager | DDG Metager | Mean |
|---|---|---|---|---|---|---|---|
| Top 1 | 0.29 | 0.20 | 0.28 | 0.50 | 0.74 | 0.47 | 0.41 |
| Top 3 | 0.26 | 0.24 | 0.25 | 0.65 | 0.71 | 0.62 | 0.46 |
| Top 10 | 0.25 | 0.24 | 0.24 | 0.64 | 0.65 | 0.62 | 0.44 |

**Table 11. Jaccard similarity of domains (US; *DDG: DuckDuckGo*)**

## DISCUSSION

Our study examined whether there are differences in the sources of the top search results between Google and alternative search engines based on popular search queries. An evaluation of root domains of the most popular sources shows only a small overlap between Google and the alternative search engines (RQ1). Overall, we found an overlap of 27% to 28% between Google and the alternatives in German results, and in the US, the overlap ranges from 24% to 25%. There is a significantly higher overlap between the alternative search engines of about 63% to 70% in German results and 62% to 65% in US results. This may be explained by all three alternative search engines using Bing's index at least in part. However, our findings show that Metager consistently has the highest overlap with Bing, going up to 78% and only overlapping as much as 64% with DuckDuckGo. The lower overlap of Google with the alternative search engines had already been shown in the studies of Agrawal et al. (2016) and Makhortykh et al. (2020). Our study provides further evidence for this.

When looking at the uniqueness of sources across all German queries, we found that the search engines returned a similar total number of sources, ranging from 2,693 to 2,841. However, for the US queries, the variety in Google was noticeably higher, with over 4,000 total sources compared to around 3,500 of the alternatives.

The most popular domain was Wikipedia, followed by sources we classified as News services (RQ2). This is likely exacerbated by selecting Google Trends as the source of our search queries, which usually includes queries related to popular news stories, sports, and celebrities. Still, it is consistent with previous studies' findings (Steiner et al., 2022) which already showed that Wikipedia and news made up the majority of search results. Furthermore, in our research, Wikipedia and news sources were the top domains across all results and the most frequent sources for each search engine individually.

When comparing the result sets of the top 10 results from Google, Bing, DuckDuckGo, and Metager, using 3,537 queries generated from Google Trends from Germany and the US, it is interesting to see that news sources are far more prevalent in German results, with social media being very infrequent. On the other hand, the US results had more social media websites. Interestingly, in the US results, YouTube was in the top 10 most popular domains in all search engines *but* Google. The same was the case for the top 50 domains in German results. This is unexpected because YouTube is a subsidiary of Google. However, this finding may be explained by the fact that we only collected organic search results, and Google might be using universal search results to display YouTube results. Another interesting difference is the greater preference of Wikipedia in Google in US results (1,892, 10,2% of all domains) compared to German results (658, 4% of all domains). Furthermore, the second and third most popular sources on Google are Instagram and Facebook for the US results, while they are not even in the top 10 of German Google sources. Finally, in terms of what is missing from Google in the US results, it is notable that Fox News is not found in the top 50 sources, while it is present in all of the alternatives.

The concentration of sources and source diversity showed a tendency for only a few root domains to make up a large share of search results. The Gini Index values of 0.73 and 0.79 in Germany and the United States, respectively, are a clear indicator (RQ3). This is consistent with findings from previous studies (Höchstötter & Lewandowski, 2009) that showed that only a few top sources dominate the search results in search engines.

Of course, our study is not without limitations. First, the selection of search queries is the most significant factor in compiling the data that was evaluated. Even though the number of queries is high and there is some diversity in the queries, the topics are almost always focused on news, celebrities, and sports, which inevitably leads to many news sources in the search results. A refined approach to selecting search queries would be appropriate for a more accurate evaluation of source diversity. For example, this could be achieved by focusing on socially controversial topics and choosing the queries accordingly.

Further limitations arise in the evaluation of source types. The classification we used in this study is very broad. A more precise classification would help make statements about the actual kinds of sources and thus to determine even more precisely which preferences and biases exist in different search engines. For example, grouping sources according to seriousness and reliability could serve as an explanation for the selection of sources to be displayed in search results. Regarding the search results collected in this study, another limitation is that only organic search results were considered. We did not consider advertisements or universal search results, although these have a strong influence on what users see on the SERP. The relevance of the search results was also not considered.

**CONCLUSION**

This study provides important insights into whether, although Google is by far the most popular search engine, the use of alternatives could benefit users. Our results show that using another or more than one search engine leads to seeing more diverse search results, allowing users to inform themselves more comprehensively. It should be noted that within each search engine's results, the concentration of sources shows that only a few top sources dominate the results, meaning whichever search engine a user chooses to use will shape what sources the information they get to see comes from.

**RESEARCH DATA**

Research data is available at: https://osf.io/nt3wv/

**ACKNOWLEDGMENTS**

This work is funded by the German Research Foundation (DFG – Deutsche Forschungsgemeinschaft; Grant No. 460676551).